\documentclass[prl,twocolumn,preprintnumbers,amsmath,amssymb]{revtex4}
\usepackage{graphicx}
\usepackage{bbm}
\usepackage{float}

\begin{document}

\preprint{}
\title{Magnetic and Electric Phase Control 
in Epitaxial EuTiO$_3$ from First Principles}
\author{Craig J. Fennie and Karin M. Rabe}
\affiliation{Department of Physics and Astronomy, Rutgers University,
        Piscataway, NJ 08854-8019}
\date{\today}

\begin{abstract}
We propose a design strategy - based on the coupling of spins, 
optical phonons, and strain - for systems in which magnetic 
(electric) phase control can be achieved by an applied 
electric (magnetic) field. Using first-principles 
density-functional theory calculations, we present a 
realization of this strategy for the magnetic perovskite 
EuTiO$_3$.
\end{abstract}


\maketitle


There is great interest in multiferroic materials in
which ferroelectric (FE) and ferromagnetic (FM) ordering
not only coexist, but in which the electrical polarization
{\bf P} and the magnetization {\bf M } are large and strongly
coupled~\cite{fiebig.review,spaldin.fiebig}. One challenge
in identifying strongly coupled FM-FE's that has received 
considerable attention in the past~\cite{nicola} is the 
scarcity of such materials in nature, as most insulators 
(a requirement for ferroelectricity) are paraelectric (PE) 
and antiferromagnetic (AFM). With the recent advances in 
first-principles density-functional methods for predicting
\cite{ederer.review} and in novel synthetic techniques
for growing new FM-FE multiferroics, the focus has now
turned to how to produce a strong coupling between the
two distinct order parameters~\cite{tokura.2006}. Recently, 
attention has focused on a mechanism in which magnetic order
itself breaks inversion symmetry~\cite{chapon.prl.06,harris.jap.06,
mostovoy.prl.06, sergienko.prb.06}. Based on this, remarkable
control of the FE state by an applied magnetic field
has been demonstrated in some rare-earth manganites
~\cite{kimura,namjung}, however, the natural scale of the spontaneous
polarization thus induced is very small, of the order
of nC/cm$^2$. Furthermore, the magnetic state appears
to be rather insensitive to an applied electric field for this
class of materials.

As a result, it is clearly advantageous to explore other 
possible mechanisms for strongly coupled multiferroism. 
A fruitful starting point for identifying such mechanisms 
is the observation, recently discussed by Tokura~\cite{tokura.2006}, 
that the basic physics of a strong {\bf M}-{\bf P}
coupling involves a competition between different ordered
states, e.g$.$ between a FM-FE state and an AFM-PE
state. In this Letter, we present a new approach for designing
a strongly coupled multiferroic in which the interplay of
spins, optical phonons, and strain leads to such a competition.

The criteria that a system must satisfy for this proposed mechanism
to be realized are as follows: (1) It must be an AFM-PE 
insulator in which at least one infrared-active (ir) phonon 
is coupled to the magnetic order, (2) the spins in the AFM 
ground state should align with the application of a magnetic 
field of modest strength, (3) this alignment should decrease 
the frequency of the spin-coupled ir-active phonon, and, (4) 
the key to our approach, the ir-active mode of interest 
must be strongly coupled to strain. Epitaxial strain can
have profound effects on the properties of thin films~\cite{karin}.
In our design strategy we use epitaxial strain to dial into 
the region of the phase diagram where a spin-phonon-driven 
destabilization of the lattice actually occurs. The FM-FE phase
thus produced is a low-lying state competing with the AFM-PE 
ground state. As a direct result of this competition, magnetic 
and electric phase control can be achieved by an applied 
electric and magnetic field respectively. As the polarization 
of the low-lying FM-FE phase results from the freezing-in of
a soft polar phonon triggered by the spin-phonon coupling, 
it is of the same order of magnitude as prototypical soft-mode 
FEs ($\mu$C/cm$^2$).
In addition, as the phase-control region 
is approached from the low strain side, the magnetocapacitance 
diverges, while on the high strain side there may be, in 
some cases, a phase boundary between the AFM-PE phase control 
region and a true equilibrium FM-FE phase.
The use of epitaxial strain to exploit the 
relatively modest spin-phonon effects displayed by many bulk 
materials provides an exciting new design strategy in the 
pursuit of strong {\bf M}-{\bf P} coupled multiferroics.

In bulk, europium titanate EuTiO$_3$ is an AFM-PE that 
crystallizes in the cubic perovskite structure (space 
group Pm$\bar{3}$m) with room temperature lattice constant
equal to that of SrTiO$_3$, $a$=3.905\AA. The Eu$^{2+}$ 
moments ($J$=$S$=7/2) order at $T_N$=5.5K~\cite{katsufuji.prb.01,
chien.prb.74,mcguire.jap.66} while neutron diffraction studies 
indicate G-type AFM order~\cite{mcguire.jap.66}.
Both diffraction and local structural probes~\cite{ravel.physica.95} 
(XANES) support EuTiO$_3$ remaining cubic at all temperatures.
Recently, Katsufuji and Takagi~\cite{katsufuji.prb.01} (KT) showed
that at the onset of AFM order, the static dielectric 
constant $\epsilon_0$ undergoes a sharp reduction of about 10\%, 
indicating a hardening of the lattice. Further, KT demonstrated 
that in an increasing magnetic field $\epsilon_0$
increases, saturating at a field large enough to fully align
the spins (approximately 1.5 T). KT argue and, as we will see,
our first-principles calculations confirm, this behavior
is due to a coupling between the spins and an ir-active
phonon of the type $\omega = \omega_{PM} +\lambda
\langle{\bf S}_i\cdot{\bf S}_j \rangle$
\cite{fennie.prl.06,baltensperger}. These bulk measurements 
show that EuTiO$_3$ satisfies the first three criteria, and
a close structural analogy to SrTiO$_3$ suggests that it 
satisfies the key fourth criterion. In the remainder of
the Letter we present a first-principles calculation of the 
ground-state epitaxial phase diagram for EuTiO$_3$ 
demonstrating that this system is indeed a realization of
our proposed mechanism for strongly coupled multiferroicity.

Within density-functional theory, the failure of the 
generalized gradient approximation (GGA) properly to 
capture the physics of strongly correlated systems is
well established. A widely accepted approach beyond 
GGA is the GGA plus Hubbard U (GGA+U) method~\cite{anisimov.jpcm.97}.
We perform first-principles density-functional calculations
using projector augmented-wave potentials within 
spin-polarized GGA+U approximation as implemented in the
{\it Vienna ab initio Simulation Package}~\cite{VASP,PAW} 
with a plane wave cutoff of 500 eV and a 6$\times$6$\times$6 
$\Gamma$-centered $k$-point mesh. The PAW potential for Eu
treated the 4f$^7$ 5s$^3$ 5p$^6$ 6s$^2$ as valence states. 
All calculations were performed with collinear spins and 
without LS-coupling. As expected for Eu$^{2+}$ which lacks
orbital degrees of freedom, inclusion of LS-coupling does 
not change the results. Values of the Eu on-site Coulomb, 
U = 6 eV, and exchange, J$_H$=1.0 eV, parameters were used
that give a reasonable account of the magnetic exchange
constants which were extracted by mapping GGA+U calculations
of the total energy for different spin configurations at T=0 
onto a classical Heisenberg model; 
$E_{spin}$=$-\sum_{ij} J_{ij}{\bf S}_i\cdot{\bf S}_j $
(note, in our notation the energy per spin bond is 2$J$).
Phonon frequencies and eigendisplacements were
calculated using the direct method where each ion was moved 
by approximately 0.01\AA. Born effective charge tensors were 
calculated by finite differences of the polarization using the
modern theory of polarization \cite{king-smith.prb.93} as 
implemented in {\sf VASP}. Before we proceed we note that 
calculations with GGA+U overestimate lattice constant $a$ by
$\sim$1$\%$. Therfore, we introduce a shift of the zero for 
$\sigma_{33}$,  the out-of-plane component of the stress, so 
that $\sigma_{33}$=0 for the cubic structure at the experimental 
lattice constant, $a_{exp}$= 3.9\AA. Thus, the correct cubic structure
is obtained at misfit strain $\eta$$\equiv$$(a_{exp}-a)/a_{exp}$=0.

\begin{figure}[b]
\includegraphics[scale=.55]{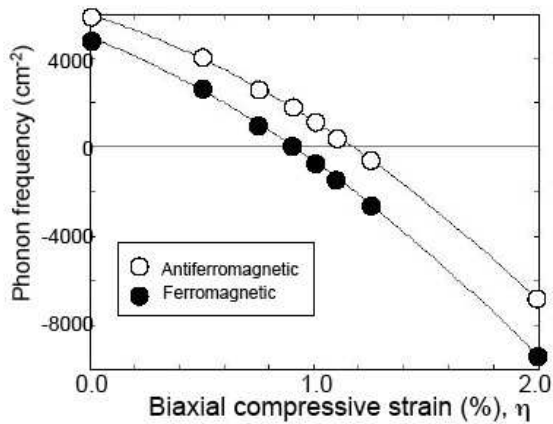}\\
\caption{\label{fig:phonons}
Soft ir-active phonon frequency squared, $\omega^2$, (cm$^{-2}$) of paraelectric
EuTiO$_3$ as a function of compressive epitaxial strain $\eta$. Space
group for $\eta$$\ne$0 is P4/mmm; the phonon is polarized along the tetragonal axis,
with symmetry label A$_{2u}$.}
\end{figure}

First, to examine the validity of a spin-phonon coupling 
mechanism in bulk EuTiO$_3$ we performed first-principles 
density-functional calculations of the ir-active 
phonons and the Born effective charges, allowing us to 
evaluate $\epsilon_0$
of a single-domain, stoichiometric, defect-free crystal. 
To elucidate the role of magnetic order on the phonons
and subsequently the dielectric response, we calculate 
the quantities just mentioned with the Eu spins in two 
different magnetic configurations: FM and G-type AFM. 
All calculations for cubic EuTiO$_3$ were carried out at
the experimental lattice constant $a_0$ = 3.90 \AA. The 
calculated magnetic exchange constants were found to be in
good agreement with experiment (shown in parentheses): 
J$_1$ = -0.013 K (-0.014 K), J$_2$ = +0.065 K (+0.037 K); 
the positive sign for J$_2$ is consistent with the unusual 
positive sign of the Curie-Weiss temperature for AFM 
EuTiO$_3$. The ground state is thus correctly found to be 
G-type AFM, with the FM state higher in energy by 0.2 meV 
per formula unit. For the static dielectric constant, we 
find $\epsilon_0$ = 350 
for EuTiO$_3$ with FM order and $\epsilon_0$ = 280 with AFM 
order.  These values compare well with the experimental 
values of 420 and 380 in fields of 1.5 T and 0 T, 
respectively~\cite{katsufuji.prb.01}. The 
spin-dependence of the dielectric response is due 
entirely to the lowest-lying ir-active phonon TO1,
which we calculate to be at 70 cm$^{-1}$ and 77 cm$^{-1}$ 
for FM and AFM respectively. This good agreement of the 
relative magnitudes of the values for $\epsilon_0$  with 
the size of the experimentally measured field-induced change
confirms KT's suggestion that the magnetocapacitive effect in
bulk EuTiO$_3$ is indeed due to spin-phonon coupling, with the
sign of the coupling $\lambda$ such that FM spin alignment 
reduces the stability of the low-frequency polar phonon.

Next, we consider the effects of epitaxial strain. In 
perovskite titanates such as SrTiO$_3$ and BaTiO$_3$, 
it is well known both experimentally and theoretically 
that epitaxial strain couples strongly to the low lying 
TO1 polar mode. For example, in SrTiO$_3$, a modest 
epitaxial strain of less than 1\% transforms the PE bulk
material into a room-temperature FE~\cite{schlom.nature.04}, 
with {\bf P} of nearly 20 $\mu$C cm$^{-2}$~\cite{antons}. 
As the calculated eigendisplacement pattern of this polar
mode for EuTiO$_3$ is virtually identical to that of the 
soft polar mode previously calculated for SrTiO$_3$ and 
BaTiO$_3$, it is natural to expect similar strain coupling, 
with the spin-phonon coupling in EuTiO$_3$ adding richness 
to the phase diagram. 

We isolate the effects of epitaxial strain, as in previous 
phenomenological and first-principles studies~\cite{pertsev,neaton,dieguez},
by imposing the epitaxial constraint on the lattice parameters 
of the infinite crystal with periodic boundary conditions, 
corresponding to zero macroscopic electric field. For 
simplicity, in the present study we consider only compressive
strain. We focus attention on the lowest-lying ir-active 
phonon TO1 which as mentioned above dominates the spin-dependent
dielectric response in cubic EuTiO$_3$ and any possible 
FE instability. In the cubic phase, TO1 is 3-fold
degenerate, where each phonon mode is polarized along one of 
the three Cartesian coordinates. Under a biaxial compressive 
strain in the $a$-$b$ plane, the lattice expands along the 
$c$-axis, conserving volume to a first approximation. This 
lowers the symmetry of EuTiO$_3$ from cubic to tetragonal, 
splitting the 3-fold degenerate TO1 mode into a two-fold E$_u$
and one-fold A$_{2u}$ degenerate mode polarized perpendicular 
and parallel to the $c$-axis respectively. Under compressive 
strain, the E$_u$ modes harden and are not relevant to our study. 

In Fig.~\ref{fig:phonons}, we show the evolution of the 
lowest-lying ir-active A$_{2u}$ phonon frequency with
biaxial compressive strain $\eta$ for FM and AFM EuTiO$_3$. 
At $\eta$=0, we have bulk 
EuTiO$_3$ where $\omega_{A_{2u}}$=$\omega_{TO1}$ = 70 cm$^{-1}$
and 77 cm$^{-1}$ for FM and AFM respectively, as we previously 
discussed. With increasing compressive strain, the A$_{2u}$ 
phonon frequency decreases for both types of  magnetic order. 
At approximately $\eta_{c1}$=0.92\%, the FM system develops
a polar instability as evidenced by the vanishing ir-active
phonon frequency. On the other hand, the AFM system remains
stable up to approximately $\eta_{c2}$=1.2\%. Thus, in the intermediate 
strain regime between $\eta$$\approx$0.92\% and $\eta$$\approx$1.2\%, 
EuTiO$_3$ has a low-lying FM-FE state in competition with a AFM-PE state, 
exactly the conditions which have been identified as facilitating
strong {\bf M}-{\bf P} coupling.

\begin{figure}[b]
\includegraphics[scale=.355]{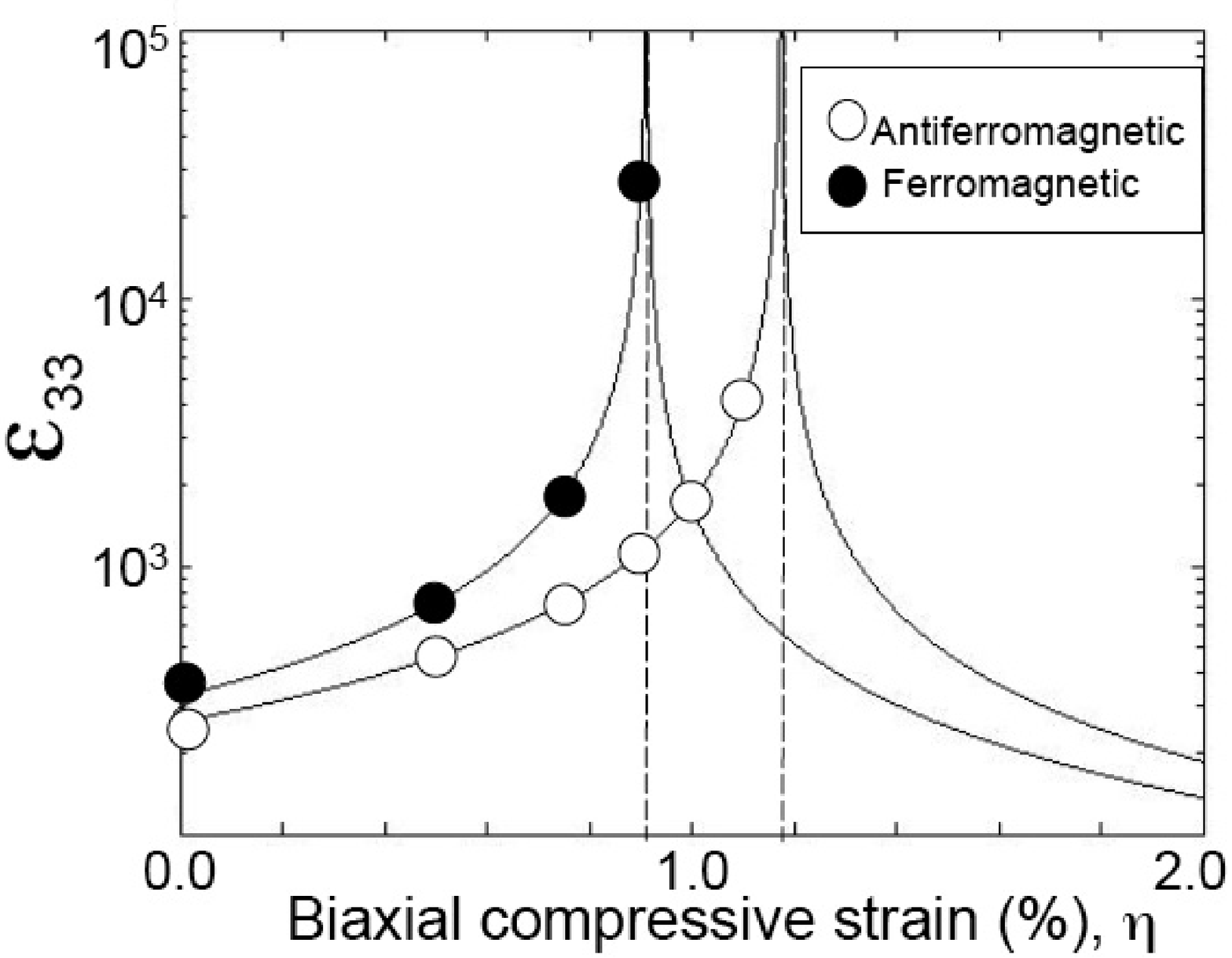}\\
\includegraphics[scale=.55]{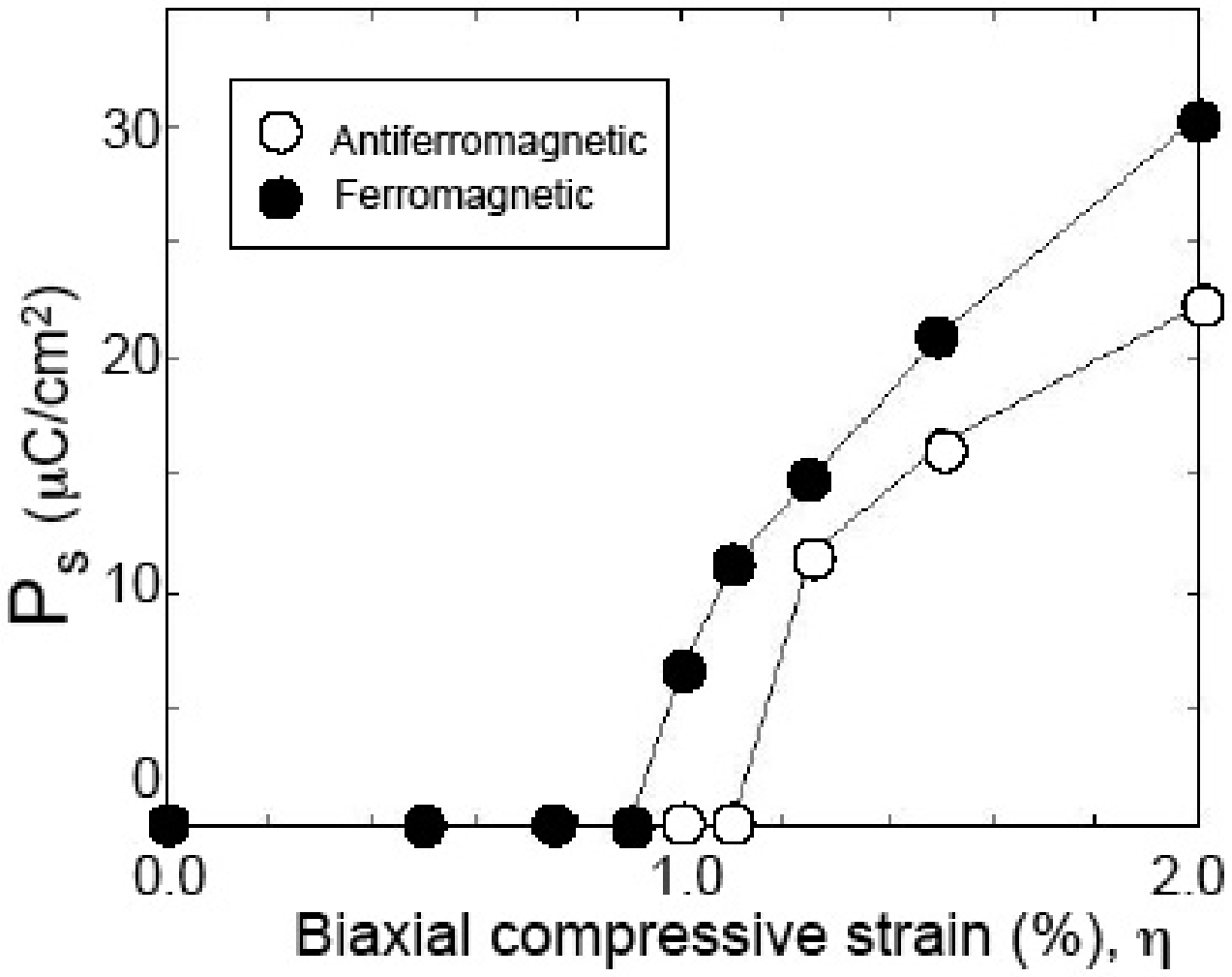}\\
\caption{\label{fig:die}
(Top) Static dielectric constant $\epsilon_{33}$ and (bottom) spontaneous polarization $P_s$
as a function of epitaxial compressive strain $\eta$. For $\epsilon_{33}$, the solid lines
are fits proportional to ($\eta$-$\eta_c$)$^{-1}$ while the dashed vertical lines indicate $\eta_c$. }
\end{figure}

It is this difference in the FE instability for the AFM and
FM orderings that leads to a giant magnetocapacitive effect 
near the critical strains, $\eta_c$. To see this, in Fig.
\ref{fig:die}(a) we show the component of the static dielectric
tensor along the $c$-axis, $\epsilon_{33}$, calculated from 
first principles~\cite{dielectric} for both AFM and FM ordering.
In each case, as the frequency of the polar mode goes to zero
at the corresponding $\eta_c$, the dielectric response diverges.
For example, at $\eta$=0.9\%, we find $\epsilon_{33}$$\approx$ 
10$^3$ for the ground state AFM-PE phase. If a magnetic field 
is applied that aligns the spins into FM ordering, $\epsilon_{33}$
increases 50-fold to that of the FM-PE phase, 
$\epsilon_{33}$$\approx$5$\times$10$^4$. Since the AFM-PE 
response remains finite as $\eta_{c1}$ is approached while the
FM-PE response diverges, the magnetocapacitance diverges in the
vicinity of the phase boundary. 

Once the critical strain is crossed for each ordering, the 
unstable polar phonon freezes in and the ground state 
crystallographic symmetry becomes $P4mm$. In Fig.~\ref{fig:die}(b),
we show the computed spontaneous polarization for the relaxed 
ground state structures as a function of epitaxial strain. A 
large {\bf P}-$\eta$ coupling is evident for both orderings.
For strains above 1.2\%, the AFM state also become FE,
but the range of strains over which the AFM-FE state is the 
ground state is extremely narrow, with a first-order transition 
to the FM-FE phase occurring at $\eta \approx$ 1.25\%. At this phase 
boundary, the polarization in the FM phase is already over 10 
$\mu$C cm$^{-2}$. Thus, we show that a previous unknown multiferroic
phase of EuTiO$_3$, with a symmetry-allowed linear magnetoelectric 
effect, can be stabilized by readily accessible epitaxial strains. 

\begin{figure}[t]
\includegraphics[scale=.3]{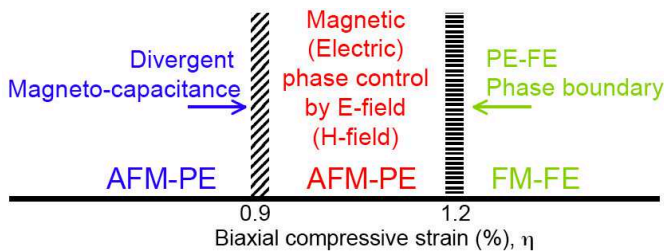}\\
\caption{\label{fig:phase}
EuTiO$_3$: Compressive epitaxial strain ($\eta$) phase diagram.}
\end{figure}

The most interesting behavior in this system is produced in the
intermediate strain region, $\eta_{c1}$ $<$ $\eta$ $<$ $\eta_{c2}$, 
under applied electric $\mathcal{E}$ or magnetic $\mathcal{H}$ fields.
We obtain a qualitative understanding of the phase diagram to 
leading order by adding the terms $-${\bf M}$\cdot\mathcal{H}$ and 
$-${\bf P}$\cdot\mathcal{E}$ to the energy, and taking {\bf M} and
{\bf P} to have their zero-field values. For example,
at $\eta$=1.0$\%$ the ground state is still the AFM-PE phase but the 
application of a magnetic field of sufficient strength to
fully align the spins, $\mathcal{O}$(1T) (the same order
as that found by KT for the bulk), induces a substantial spontaneous
polarization {\bf P}$\approx$10$\mu$C cm$^{-2}$, by driving the 
system into the FM-FE phase. Notably, {\bf P} is several 
orders of magnitude greater than the polarization of
all previously known FEs whose origin is through a coupling to spins. 
In addition, at  $\eta$=1.0$\%$ the application of an electric field,
$\mathcal{O}$(10$^5$V cm$^{-1}$) (small for thin films) favors the 
polar FE phase over that of the non-polar PE phase, inducing a
magnetization of 7$\mu_b$ by driving the system into the FM-FE phase.
We summarize the results for compressive strain in Fig.~\ref{fig:phase}.

From these estimates, both magnetic control of the electric phase 
and electric control of the magnetic phase in epitaxially strained 
EuTiO$_3$ should be easily attainable experimentally. However, the
symmetry of the PE-paramagnetic structure determines 
the lowest order spin-phonon coupling (the origin of the discussed 
effects) as a bi-quadratic effect $\mathcal{F}_{int} 
\propto {\bf M}^2 {\bf P}^2 $. 
So, while an applied magnetic field can turn the polarization on 
from zero, there is no preferred direction and the magnetic field 
does not act to reversibly switch the polarization 
between symmetry-related orientations. At present, such switching 
has only been demonstrated in multiferroics with minute 
polarizations~\cite{namjung,kimura}. Similarly, an applied electric 
field cannot be used to reversibly switch between different 
magnetization states~\cite{fiebig.nature}, though it can turn a 
substantial magnetization on from zero.

In further analogy to SrTiO$_3$, effects similar to those descibed
above in EuTiO$_3$
are expected not only for compressive, but also for tensile 
epitaxial strain; first-principles investigations are in progress.
The recent success in fabricating high-quality thin films of 
EuTiO$_3$ with zero epitaxial strain~\cite{suzuki} is encouraging
for experimental realization of the predicted effects. While we 
have focused on the example of EuTiO$_3$, we expect that additional
materials with the necessary characteristics could be identified 
and their epitaxial phase diagrams explored.

In summary we have presented a realization of a new mechanism 
for strong coupling between magnetic and ferroelectric ordering.
The required material characteristics are a spin-phonon coupling 
through which FM spin alignment softens a low-frequency polar mode 
that is strongly coupled to epitaxial strain. The predicted 
competition between an AFM-PE phase and a FM-FE phase allows 
magnetic phase control with an applied electric field, and electric
phase control with an applied magnetic field, with modest critical
fields. In addition to being a promising mechanism by which such 
phase-control can be achieved, we anticipate that epitaxial 
stabilization of a FM-FE ground state - such as that which occurs
in EuTiO$_3$ above strains $\approx$ 1.25$\%$ - over the bulk 
AFM-PE phase may also prove to be a useful avenue to the 
identification of new FM insulators suitable for spintronics applications.

We thank D. G. Schlom and D. Vanderbilt for useful discussions. This
work was supported by NSF DMR-05-07146. CJF acknowledges support from Bell
Labs and Rutgers University.



\newpage


\begin{thebibliography}{10}
\bibitem{dielectric}
In polar dielectrics, the phonon response dominates $\epsilon_0$, 
and the electric contribution, $\epsilon_{\infty}$, can be ignored. 
Then  $\epsilon_0 = \sum_{m} {\Omega_{p,m}^2 / 
\omega_{m}^2}$ where the individual terms are the contribution from each 
ir-active phonon. $\Omega_{p,m}$ and $\omega_{m}$ are the effective 
plasma frequency and ir phonon frequency for mode $m$, 
respectively.  $\Omega_{p}$ was determined from 
calculations of the Born effective charge tensors
and the real-space eigendisplacements of the phonons.
\end{thebibliography}
\end{document}